\documentclass[showpacs]{revtex4}
\usepackage{graphicx}
\usepackage{amsmath}
\begin{document}
\title{Nonperturbative Quantization of the 
Cylindrically Symmetric Strongly Oscillating 
Field}
\author{V. Dzhunushaliev}
\email{dzhun@hotmail.kg} 
\affiliation{Phys. Dept., Kyrgyz-Russian Slavic University, 
Bishkek, 720000, Kyrgyz Republic}

\begin{abstract}
A recent investigation of SU(2) Yang-Mills theory found 
several classical solutions with bad behaviour at 
infinity : one of the potential components oscillated 
and another tended to infinity. In this paper we apply
an idea due to Heisenberg
about the quantization of strongly interacting nonlinear 
fields to these classical singular solutions. We find that 
this quantization procedure eliminates the bad long distance
features while retaining the interesting short distance
aspects of these solutions.
\end{abstract}
\maketitle

\section{Introduction}

In quantum chromodynamics it is postulated that 
flux tubes filled with SU(3) color field 
are stretched between quark and antiquark giving rise
to confinement. Neglecting the cross section of
the flux tube gives an effective string connecting the
quark and antiquark. There are various approaches to describing
such strings: Veneziano's amplitude \cite{venez}, 
Polyakov's string \cite{polyakov}, 
the Nielsen-Olesen string \cite{nielsen}, 
the Kleinert-Chervyakov string \cite{kleinert}
{\it etc.}. In Refs. \cite{obukhov,dzh1} several
cylindrically symmetric, SU(2) gauge field solutions
were investigated, but these solutions either had
a singularity \cite{obukhov} or had bad
asymptotic behaviour \cite{dzh1}. The flux tube
solution of Ref. \cite{dzh1} had one gauge field
component which grew linearly at infinity, while
another component oscillated with increasing frequency.
In this paper it was shown that 
the bad asymptotic behaviour was improved after
applying a quantization procedure first employed by
Heisenberg for strongly interacting, non-linear fields.
\par
The physical idea presented here is very simple:
\begin{itemize}
\item
quantum fluctuations ``smooth'' the gauge
fields which oscillate strongly,
\item
to a first approximation the quantization of the 
strong oscillating field leads to an averaging over all
classical solutions.
\end{itemize}

\section{Classical SU(2) solution}

We briefly review the classical, cylindrically symmetric,
SU(2) Yang-Mills theory solutions\cite{dzh1}. 
Consider the following ansatz for SU(2)
Yang-Mills theory
\begin{eqnarray}
A^a_t & = & f(\rho ), \; a=1
\label{16:1}\\
A^a_z & = & v(\rho ), \; a=2
\label{16:2}
\end{eqnarray}
here we use the cylindrical coordinate system $z, \rho , \varphi$ 
and the SU(2) color index $a=1,2,3$. With this
ansatz the Yang - Mills equations
\begin{equation}
D_\mu F^{\mu\nu} = 
\frac{1}{\sqrt{-g}}\partial _\mu 
\left (\sqrt {-g} F^{a\mu\nu}
\right ) + \varepsilon^{abc} F^{b\mu\nu} A^c_{\mu} = 0,
\label{16:3}
\end{equation}
become
\begin{eqnarray}
f'' + \frac{f'}{\rho} & = & fv^2,
\label{17:1}\\
v'' + \frac{v'}{\rho} & = & -vf^2,
\label{17:2}
\end{eqnarray}
primes denote a derivative with respect to 
$\rho$; $\varepsilon^{abc}$ is the structure 
constant of the SU(2) gauge group. 
The asymptotic behaviour of the ansatz functions $f, v$ 
and the energy density ${\cal E}$ can be given as
\begin{eqnarray}
f & \approx & \frac{2}{\rho _0}\left [x + 
\frac{\cos \left (2x^2 + 2\phi\right )}
{16x^3} \right ] ,
\label{21:1}\\
v & \approx & \frac{\sqrt{2}}{\rho_0} \frac{\sin 
\left (x^2 + \phi\right )}{x} ,
\label{21:2}\\
{\cal E} & \propto & {f'}^2 + {v'}^2 + f^2v^2 \approx const,
\label{21:3}
\end{eqnarray}
where $x=\rho /\rho _0$ is a dimensionless
radius, and $\rho _0, \phi$
are  constants. A numerical investigation of
Eqs. (\ref{17:1}-\ref{17:2}) had been carried
out in Ref. \cite{dzh1}.
This inquiry showed that $A^1_t=f(\rho)$ was
a confining potential while $A^2_z=v(\rho)$
was a strongly oscillating potential . 

\section{Quantization}

\subsection{Heisenberg Quantization of Strongly Interacting Fields}

Heisenberg's idea of the 
quantization of strongly interacting fields 
has its origin in the fact that the n-point Green's
functions can be found from some infinite set of differential equations
derived from the field equations for the field operators. As an example
we present Heisenberg's method of quantization for a nonlinear 
spinor field \cite{hs1} \cite{hs3}.
\par
The Heisenberg equation for a quantum spinor field is
\begin{equation}
\label{h1}
\gamma ^{\mu} \partial _{\mu} {\hat \psi} (x) -
l^2_0 \Im [{\hat \psi}(x) ({\hat {\bar \psi}}(x)
{\hat \psi}(x)) ] = 0
\end{equation}
where $\gamma ^{\mu}$ are Dirac matrices;
${\hat \psi} (x), {\hat {\bar \psi}}(x)$ are
the operator of the spinor field and its adjoint respectively;
the nonlinear term 
$\Im [{\hat \psi} ({\hat {\bar \psi}}
{\hat \psi)} ] = {\hat \psi} ({\hat {\bar \psi}}
{\hat \psi})$ or ${\hat \psi} \gamma ^5
({\hat {\bar \psi}} \gamma ^5{\hat \psi})$ or
${\hat \psi} \gamma ^{\mu}
({\hat {\bar \psi}} \gamma _{\mu} {\hat \psi})$ or
${\hat \psi} \gamma ^{\mu} \gamma
^5 ({\hat {\bar \psi}} \gamma _{\mu} \gamma ^5{\hat \psi} )$; 
$l_0$ is some constant. 
Heisenberg emphasized that the 2-point Green's function,
$G_2 (x_2, x_1)$, in this theory differs strongly from the propagator in
a linear theory. This difference lies in its behaviour on the light
cone : in the nonlinear theory $G_2 (x_2 , x_1)$ oscillates strongly on
the light cone in contrast to the propagator of the linear theory
which has a $\delta$-like singularity. Heisenberg introduced the
$\tau$ functions as
\begin{equation}
\label{h2}
\tau (x_1 x_2 ... | y_1 y_2 ...) = \langle 0 |
T[{\hat \psi} (x_1) {\hat \psi} (x_2) ...
{\hat {\bar \psi }} (y_1) {\hat {\bar \psi }} (y_2) ...] |
\Phi \rangle
\end{equation}
where $T$ is the time ordering operator; $| \Phi \rangle$ is a state for
the system described by Eq. (\ref{h1}). Applying Heisenberg's
equation (\ref{h1}) to (\ref{h2}) we can obtain the following infinite
system of equations for various $\tau $'s
\begin{eqnarray}
\label{h3}
l_0^{-2} \gamma ^{\mu} _{(r)} \frac{\partial}{\partial x^{\mu} _{(r)}}
&&\tau (x_1 ...x_n |y_1 ... y_n ) = \Im [ \tau (x_1 ... x_n x_r |
y_1 ... y_n y_r)] + \nonumber \\
&&\delta (x_r -y_1) \tau ( x_1 ... x_{r-1} x_{r+1} ... x_n |
y_2 ... y_{r-1} y_{r+1} ... y_n ) + \nonumber \\
&&\delta (x_r - y_2) \tau (x_1 ... x_{r-1} x_{r+1} ... x_n |
y_1 y_2 ... y_{r-1} y_{r+1} ... y_n ) + ...
\end{eqnarray}
Eq. \eqref{h3} represents one of an infinite set of coupled equations
which relate various order of the $\tau$
functions to one another. In fact Heisenberg 
showed that differential equations for the field operators are
equivalent to some infinite set of differential equations for
Green functions (for a small coupling constant this is the
Dyson-Schwinger system of equations).
\par
We know that the standard Feynman diagram 
technique for dealing with field 
theories via an expansion in terms of a small parameter does 
not work for strongly coupled, nonlinear fields. Heisenberg 
used the above procedure to study the 
nonlinear Dirac equation. Our basic goal in 
this paper is to apply the Heisenberg method to a 
flux tube-like solution of classical SU(2) 
Yang-Mills theory. Under certain assumptions about 
this quantization method we will find that
the singular, asymptotic behaviour of the classical Yang-Mills
gauge field is ``smoothed'' out when the Heisenberg 
quantization technique is applied. 

\subsection{Quantization of the strongly oscillating classical solutions}

In order to simplify
Heisenberg's quantization method to the present nonlinear equations
we make the following assumptions \cite{dzh2}:
\begin{enumerate}
\item
The physical degrees of freedom relevant for studying 
the above-mentioned classical solution 
are given entirely by the two ansatz 
functions $f ,v$ appearing in Eqs. \eqref{17:1}, \eqref{17:2}. 
No other degrees of freedom will arise through 
the quantization process.
\item
From Eqs. \eqref{21:1}, \eqref{21:2} we see that one function
$f(r)$ is a smoothly varying function for large
$x$, while another function, $v(r)$, is strongly oscillating.
Thus we take $f(r)$ to be almost a
classical degree of freedom while
$v(r)$ is treated as a
fully quantum mechanical degree of freedom.
Naively one might expect that
only the behaviour of second function
would change while first function stayed
the same. However since both functions are interrelated
due to the nonlinear nature of the equations of motion
we find that both functions are modified.
\end{enumerate}
To begin applying Heisenberg's quantization scheme to this Yang-Mills
system we replace the ansatz functions by operators ${\hat f} (\rho) ,
{\hat v} (\rho )$ (here we follow Refs. \cite{dzh2}):
\begin{eqnarray}
{\hat f}'' + {{\hat f}' \over x} &=& {\hat f} {\hat v}^2 ,
\label{s9:1} \\
{\hat v}'' + {{\hat v}' \over x} &=& -{\hat v} {\hat f}^2 .
\label{s9:2}
\end{eqnarray}
Taking into account assumption (2) we let 
${\hat f} \rightarrow f$ become just
a classical function again, and replace 
$\langle v \rangle = \langle \Phi |{\hat v} | \Phi \rangle$, 
$\langle v^2 \rangle = \langle \Phi |{\hat v} | \Phi \rangle$ 
(here $\Phi$ is a quantum state) 
in Eq. ({\ref{s9:1}-\ref{s9:2}) by its expectation value
\begin{eqnarray}
f'' + {f' \over x} &=& f \langle v^2 \rangle ,
\label{s10:1} \\
{\langle v \rangle }'' + 
\frac{{\langle v \rangle }'}{x} &=& 
- \langle v \rangle f^2 .
\label{s10:2}
\end{eqnarray}
In Ref.\cite{dzh2} a new equation for 
$\langle v^2 \rangle$ was found, and some assumptions 
were made for cutting off the infinite set of equations 
for $\langle {v'}^2 \rangle$, $\langle v^4 \rangle$ {\it etc.}
Here we shall proceed differently.

\section{Averaging over all classical solutions}

Our calculations presented below are based on the following
assumption 
\begin{equation}
\label{av-1}
  \langle A^a_\mu \rangle = \int  A^a_\mu e^{i S
  \left [  A^a_\mu \right ]}D A^a_\mu
  \approx
  \int \left (\tilde A^a_\mu \right )_\phi 
  e^{i S  \left [ \left (\tilde A^a_\mu
  \right )_\phi \right ]}
  D \left (\tilde A^a_\mu \right )_\phi = 
  \sum _{\substack{\text{over all} \\ 
  \text{classical solutions}}}
  \left (
  \tilde {A^a_\mu}
  \right )
  _\phi p_\phi
\end{equation}
where $A^a_\mu$ is the gauge potential, $\tilde A^a_\mu$ 
is the classical (possibly singular) solution of the 
Yang - Mills equations numbered by a parameter $\phi$, 
$p_\phi$ is the probability for a given 
classical solution. For the single solution we have
well known expression
\begin{equation}
  \int e^{iS[\Phi]} D\Phi \approx A e^{iS[\Phi_{cl}]}
  \label{av-2}
\end{equation}
where $A$ is the normalization constant and consequently
\begin{equation}
  \int \Phi e^{iS[\Phi]} D\Phi \approx \Phi_{cl} .
  \label{av-3a}
\end{equation}
In our case $\Phi$ is the gauge potential $A^a_\mu$.
\par
It is natural to assume that in our case the classical
solutions with asymptotic behaviour (\ref{21:1}-\ref{21:2}) 
have (in the first approximation) 
identical probability $p_\phi \approx const$. Therefore
\begin{eqnarray}
  \langle v \rangle \approx 
  \frac{1}{2\pi} \int\limits^{2\pi}_0 v_{cl} d\phi = 
  \frac{1}{2\pi\rho_0} 
  \int\limits^{2\pi}_0 \frac{\sqrt 2}{x}
  \sin(x^2 + \phi) d\phi  = 0 ,
  \label{av-3}\\
  \langle v^2 \rangle \approx 
  \frac{1}{2\pi} \int\limits^{2\pi}_0 v^2_{cl} d\phi = 
  \frac{1}{2\pi\rho^2_0}
  \int\limits^{2\pi}_0
  \frac{2}{x^2}   \sin^2 (x^2 + \phi) d\phi =
  \frac{1}{\rho^2} .
  \label{av-4}
\end{eqnarray}
here $v_{cl}$ is the function from the Eq.(\ref{21:2}). 
It is necessary to note that in this quantization 
model the quantity $\rho_0$ remains classical. 
Now we can substitute Eqs. (\ref{av-3}), (\ref{av-4})
into Eqs. (\ref{s10:1}-\ref{s10:2}). 
After this Eq.(\ref{s10:1}) is satisfied identically 
and Eq.(\ref{s10:2}) has the form
\begin{equation}
  f'' + \frac{f'}{\rho} = \frac{f}{\rho^2}
  \label{av-5}
\end{equation}
with the solution
\begin{equation}
  f = \frac{f_0}{\rho}
  \label{av-6}
\end{equation}
where $f_0$ is some constant. We should note that there is 
another solution $f = K\rho$ but this solution is not interesting 
for us. The expression (\ref{av-3}), 
(\ref{av-4}) and (\ref{av-6}) gives us the following important 
result: \textbf{\textit{the quantum fluctuations
of the strongly oscillating field leads to the improvement
of the bad asymptotic behaviour of these fields.}}
This means that after quantization the linearly growing 
and strongly oscillating components of the gauge potential 
become functions with good asymptotic behaviour. 
The cross section of this cylindrically symmetric
tube can be presented as in Fig.\ref{fig1}.
\begin{figure}
\setlength{\fboxrule}{1pt}
\centerline{\framebox{
\includegraphics[width=8cm,height=5cm]{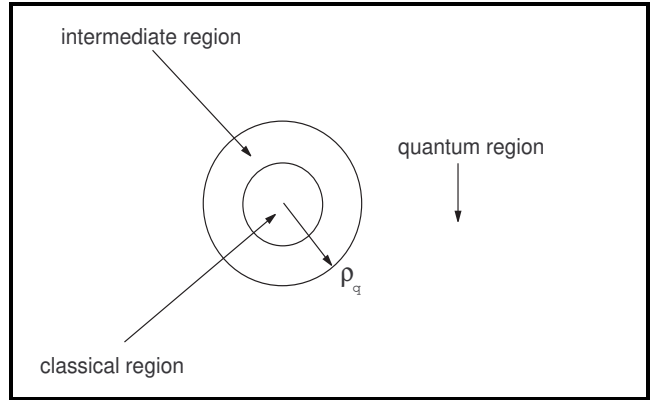}
}}
\vspace{5mm}
\caption{Cross section of the cylindrically symmetric tube. 
The field in the classical region is described
by the classical Yang - Mills equations 
\eqref{17:1}-\eqref{17:2}. The field in the
quantum region is described by the quantum Yang -
Mills equations \eqref{s10:1}-\eqref{s10:2}.}
\label{fig1}
\end{figure}
In the quantum region we have the following color fields
\begin{eqnarray}
  \langle E^1_r \rangle = f' & \approx & -\frac{f_0}{\rho^2} ,
  \label{av-7}\\
  \langle E^3_z \rangle = f\langle v \rangle & \approx & 0 ,
  \label{av-8}\\
  \langle H^2_\varphi \rangle = \langle v' \rangle
  & \approx & 0
\end{eqnarray}
This means that we have a flux tube concentrated in the 
classical region. 

\section{Energy density}

To calculate the energy density 
$\varepsilon = (E^a_\mu)^2 + (H^a_\mu)^2$ we need
to find
\begin{eqnarray}
   \langle (E^1_r)^2 \rangle = {f'}^2 & \approx & 
   \frac{f_0^2}{\rho^4} ,
\label{en-1}\\
   \langle (E^3_z)^2 \rangle = f^2 \langle v^2 \rangle 
   & \approx & \frac{f_0^2}{\rho^4} ,
\label{en-2}\\
   \langle (H^2_\varphi)^2 \rangle = 
   \langle {v'}^2 \rangle & - & ? 
\label{en-3}
\end{eqnarray}
here $1,2,3$ are color indices. 
We see that $\langle (E^1_r)^2 \rangle$ and 
$\langle (E^3_z)^2 \rangle$ have good asymptotic 
behaviour. To calculate $\langle (H^2_\varphi)^2 \rangle$ 
we use the following equation 
\begin{equation}
  \langle {v'}^2 \rangle ' = 2\langle v' v'' \rangle = 
  -2\frac{\langle {v'}^2 \rangle}{\rho} - 
  \langle v^2 \rangle ' f^2
  \label{en-4}
\end{equation}
which has the following solution 
\begin{equation}
  \langle {v'}^2 \rangle = -\frac{f_0^2}{\rho^4} .
  \label{en-5}
\end{equation}
One can see immediately that this result is incorrect.
This probably is connected with the rough approximation
of Eqs. (\ref{av-3}, \ref{av-4} and \ref{av-6})
for $f$, $\langle v \rangle$ and $\langle v^2 \rangle$.
One possible remedy would be to try a more realistic
approximation, in which $\rho_0$ is also a
quantized quantity. 

\section{Defining the radius of the quantum region}

To define a radius, $\rho_q$, below which one must begin
using a quantum description of the gauge field 
we will compare the mean square of the gauge field 
$\sqrt{\langle (E^2_z)^2 \rangle}$ (in the quantum region) 
with the amplitude of the classical field $(E^2_z)_{cl}$. 
If these values are of the same order this means that 
the fluctuations 
\textbf{\textit{of the quantum gauge field 
are so large that they ``smooth'' the strong
oscillations of the classical gauge field}}. 
\par 
As $\langle v \rangle \approx 0$ then 
$\langle \Delta v^2 \rangle \approx \langle v^2 \rangle$ 
and we have 
\begin{equation}
  \sqrt{\langle (E^2_z)_q^2 \rangle} = 
  \sqrt{f^2\langle v^2 \rangle} \approx 
  \frac{f_0}{\rho ^2_q} ,
  \label{qu-1}
\end{equation}
$\rho _q$ is the radius of the quantum region 
(see Fig.\ref{fig1}) and the index (q) indicates
that this is the beginning of the quantum region.
For the classical field, $E^2_z$, we can write 
\begin{equation}
  \left (E^2_z \right )_{cl} = 
  fv \approx \frac{1}{\rho ^2_0} .
  \label{qu-2}
\end{equation}
Comparing these two expressions
\begin{equation}
\sqrt{\langle (E^2_z)^2 \rangle } = 
\left ( E^2_z \right )_{cl} ,
\label{qu-3}
\end{equation}
we define the radius 
\begin{equation}
  \rho _q = \rho _0 \sqrt{f_0}
  \label{qu-4}
\end{equation}

\section{Conclusions}

The chief point in this paper is that 
a quantum field theory with a strong coupling constant 
can have nonlocal objects ({\it e.g.} flux tube solutions). 
This allows us to conclude that 
\textit{\textbf{not every quantum object
can be simply described as a cloud of quanta.}}
\par 
Another consequence is that the flux tube can have a complicated 
structure: near the tube axis the SU(2) gauge field 
behave classically, but at infinity their behaviour is
modified by quantum mechanics. The physical reason for this 
is that quantum fluctuations smooth the strongly oscillating 
fields in such a manner that at infinity the bad long
distance behaviour is improved. 
\par 
We note that these results can be easily verified 
by numerical simulations on the lattice as the fields 
$f$ and $v$ are 1-dimensional and the Lagrangian has the
simple form ${\cal L} = {f'}^2 - {v'}^2 + f^2 v^2$. 

\section{Acknowledgment}

I am grateful for financial support from the ICTP (grant KR-154) 
Viktor Gurovich and Doug Singleton for the fruitful discussion.

\end{document}